\documentclass[aps,prd,groupedaddress,nofootinbib,floatfix]{revtex4}

\usepackage{graphicx,amsmath}
\usepackage{yfonts}
\usepackage{ulem}
\usepackage{float}
\DeclareGraphicsExtensions{.eps, .eps, .jpg, .png}

\def\e{{\mathrm {e}}}

\begin{document}




\title{Reviving chaotic inflation with fermion production: a supergravity model}



\author{Michael Roberts}

\email[]{mroberts@physics.umass.edu}
\author{Lorenzo Sorbo}

\email[]{sorbo@physics.umass.edu}

\affiliation{Amherst Center for Fundamental Interactions, Department of Physics, University of Massachusetts, Amherst, MA 01003, U.S.A.}


\date{\today}

\begin{abstract}

Processes of particle production during inflation can increase the amplitude of the scalar metric perturbations. We show that such a mechanism can naturally arise in supergravity models where an axion-like field, whose potential is generated by monodromy, drives large field inflation. In this class of models one generally expects instanton-like corrections to the superpotential.
We show, by deriving the equations of motion in models of supergravity with a stabilizer, that such corrections generate an interaction between the inflaton and its superpartner. This inflaton-inflatino interaction term is rapidly oscillating, and can lead to copious production of fermions during inflation, filling the Fermi sphere up to momenta much larger than the Hubble parameter. In their turn, those fermions source inflaton fluctuations, increasing their amplitude, and effectively lowering the tensor-to-scalar ratio for the model, as discussed in~\cite{Adshead:2018oaa,Adshead:2019aac}. This allows, in particular, to bring the model where the inflaton potential is quadratic (plus negligibly small instanton corrections) to agree with all existing observations.

\end{abstract}


\maketitle


\section{Introduction}%
\label{sec:intro}

Cosmological observations restrict the space of viable inflationary models in various directions. The measurement of the spectral index gives $n_s-1\simeq -1/30$ with a $\sim 10\%$ uncertainty, with no appreciable running. This, together with constraints on nongaussianities (the parameter $f_{NL}$ is about four orders of magnitude smaller than its value in a fully nongaussian distribution), and with the fact that isocurvature modes are below the $5\%$ level~\cite{Akrami:2018odb}, severely constrains non-vanilla models of inflation. However, the class of models of inflation that arguably taste the most like vanilla, those described by a monomial potential $V\propto\varphi^n$, are either ruled out or under significant pressure from the constraints on the tensor-to-scalar ratio $r\lesssim .06$~\cite{Ade:2018gkx}. In particular, the simplest choice for a monomial, the chaotic inflation with quadratic potential, is ruled out at the $\sim 4 \sigma$ level.

Scenarios where particle production occurs during inflation can allow to decouple inflationary observables from the shape of the potential. In these scenarios, the rolling inflaton  provides the energy necessary for the generation of quanta of a secondary field whose presence can affect the spectra of scalar and tensor perturbations. While a single event of particle production~\cite{Chung:1999ve} can lead to features in the power spectra, a continuous process can generate an additional quasi-scale invariant component for the  spectrum of scalar perturbations, or even provide a channel for the dissipation of the inflaton's energy that can lead to inflation even if the potential does not satisfy the slow-roll conditions~\cite{Berera:1995ie,Green:2009ds,Anber:2009ua,Adshead:2018oaa}.

In this paper we will argue that particle production can bring the model of chaotic inflation $V(\varphi)=\frac{\mu^2}{2}\varphi^2$ (plus, as we will see, corrections that we will require to be negligible) to agree with all constraints from observations. This possibility was already considered  in~\cite{Abolhasani:2019lwu,Abolhasani:2020xcg}, that discussed a system where an auxiliary scalar $\chi$ gets an oscillating mass-squared through a coupling to the inflaton, leading to periodic production of quanta of $\chi$. Remarkably, in our work we will see that one can resurrect chaotic inflation by simply embedding it in a supersymmetric setting, and including a small, instanton like correction to the superpotential. These ingredients - monomial inflation with small instanton corrections in supersymmetry - are expected in supersymmetric models where the inflaton is an axion-like degree of freedom whose potential is generated by monodromy~\cite{Silverstein:2008sg,McAllister:2008hb}, see~\cite{Pajer:2013fsa} for a review. In particular, the quadratic form of the inflaton potential is generated in the axion--four-form system of~\cite{Kaloper:2008qs,Kaloper:2008fb,Kaloper:2011jz,Marchesano:2014mla}. Note also that~\cite{Kaloper:2016fbr,DAmico:2017cda} have shown that this axion-four form system can be brought to agree with observations by the inclusion of higher dimensional operators that flatten the potential at large field values, similarly to the effect~\cite{Silverstein:2008sg,McAllister:2008hb,Dong:2010in,McAllister:2014mpa} observed in string theory constructions. In this work we will assume, however, that such flattening does not occur for the observationally relevant range of field values.

Going into the specifics of our scenario, we will consider a supersymmetric model where $\Phi$ is a superfield whose imaginary component $\varphi$ gives the axion-like inflaton, whose potential is generated by monodromy. We will show that the addition of an instanton-like $\sim e^{-\Phi/F}$ component to the superpotential, where  $F\ll M_P$ is the axion constant, leads to a coupling of the inflaton to its fermionic partner, the inflatino, that can be written in the form $\sim \frac{1}{F}\bar\psi \left(\gamma^\mu\gamma^5\partial_\mu\varphi\right)\psi$. The rolling inflaton thus provides a time-dependent contribution to the fermionic Lagrangian, leading to the generation of quanta of $\psi$~\cite{Adshead:2015jza,Adshead:2015kza} with momenta up to $\sim \dot\phi/F$, that can be much larger than the Hubble parameter, leading to a large density $\gg H^3$ of fermions during inflation. The quanta of $\psi$, in their turn, source fluctuations of the inflaton. The phenomenology associated to this fermion-inflaton system has been studied in~\cite{Adshead:2018oaa,Adshead:2019aac}. In the first of those papers it was shown that there is a regime where the inflaton fluctuations sourced by the produced fermions dominate over the standard ones  originating from the amplification of the inflaton's vacuum fluctuations. Remarkably, and in contrast with the case in which gauge fields are amplified by the rolling axion-like inflaton~\cite{Barnaby:2010vf}, the statistics of the inflaton perturbations is quasi-gaussian, and in agreement with observations. Also, as we will see, the spectral index in this model turns out to be the same as in standard chaotic inflation, and therefore agrees with observations.

The existence of an additional component of scalar perturbations increases, for fixed values of the parameters, the amplitude of the power spectrum $P_\zeta$. If such a component is sizable, therefore,  we must lower the energy scale of inflation in order to fit the observed value of $P_\zeta$. This has the consequence of lowering the amplitude of the tensor perturbations (in~\cite{Adshead:2019aac} it was shown that the fermions do not source significantly the tensor modes), and of bringing the model to agree with the current constraints on the tensor-to-scalar ratio $r$. This is one of the main results of our paper.  

As a warm-up, in Section~\ref{sec:susy} we will consider a globally supersymmetric model, with a superpotential $W\propto \Phi^2$, that can lead to chaotic inflation, with a small contribution $\sim e^{-\Phi/F}$. While we work directly in the regime of supersymmetry with a single chiral superfield, it is worth noticing that this same construction can be realized~\cite{Groh:2012tf} by supersymmetrizing the axion--four-form system of~\cite{Kaloper:2008qs,Kaloper:2008fb,Kaloper:2011jz}. Mapping the resulting fermionic Lagrangian to that studied in~\cite{Adshead:2018oaa}, and imposing theoretical as well as observational constraints, we find that this scenario can agree with observations, leading in particular to a tensor-to-scalar ratio that can be as small as  $r\simeq .007$, about a factor $8$ below the current bound.

Given that the inflaton has Planckian excursions, however, the assumption of global supersymmetry is not appropriate, and one has to go to the full supergravity description. We perform such an analysis in Section~\ref{sec:sugra}. We consider models that are free from the $\eta$-problem~\cite{Dine:1995uk} by choosing a K\"ahler potential that depends only on the combination~\cite{Gaillard:1995az,Kawasaki:2000yn} $\Phi+\bar\Phi$. In order to design a potential that is dominated by the quadratic term and whose flatness at large values of the inflaton is not spoiled by the supergravity correction, we consider models of inflation with a stabilizer~\cite{Kawasaki:2000yn,Kallosh:2010ug,Kallosh:2010xz} superfield $S$. Since this system features two superfields $\Phi$ and $S$, we must diagonalize the dynamics of two fermions, the inflatino and the stabilizerino, that is given in general terms in~\cite{Kallosh:2000ve}. We do so by generalizing the analysis of~\cite{Nilles:2001ry,Nilles:2001fg} to the case where the system contains a pseudoscalar component, but with the simplifying assumption that, thanks to the presence of the stabilizer, the superpotential vanishes on shell. To our knowledge, such a calculation is new in the literature, and is our other main result.

After diagonalizing the fermions, and in the regime where the inflaton potential has  small oscillations superimposed to a large monomial component, we find that the dynamics of this system is identical -- up to simple redefinitions of parameters -- to the globally supersymmetric one. Thus the parameter space contains a viable region where the scalar potential is essentially quadratic, but $r$ can be as small as a factor $\sim$ 8 below the current constraints, also in the case in which the model is embedded in supergravity.

\section{Fermion production during inflation, and the amplitude of tensor-to-scalar ratio}%
\label{sec:general}

Let us start by reviewing the results of~\cite{Adshead:2018oaa,Adshead:2019aac}. Those papers contain the study of a system consisting of  a pseudoscalar inflaton $\varphi$ with arbitrary potential $V(\varphi)$ generated by the breaking of the shift symmetry $\varphi\to\varphi+$constant, along with a fermion $Y$ of mass $m_\psi$. Including the shift symmetric coupling of lowest dimensionality of the inflaton to $Y$, the fermionic component of the Lagrangian takes the form
\begin{align}\label{eq:lagrY}
{\cal L}_{Y}=\bar{Y}\left[i\gamma^\mu\partial_\mu-m_\psi-\frac{1}{f}\gamma^\mu\gamma^5\partial_\mu\varphi\right] Y\,,
\end{align}
where $f$ is a constant with the dimensions of a mass. It is convenient to define a new fermion field $\psi$, related to $Y$ by
\begin{align}
\psi=e^{i\gamma^5\varphi/f}Y\,,
\end{align}
in terms of which the fermionic Lagrangian reads
\begin{align}\label{eq:mainferm}
{\cal L}_\psi=\bar\psi\left\{i\gamma^\mu\partial_\mu-m_\psi\,\left[\cos\left(\frac{2\varphi}{f}\right)-i\gamma^5\sin\left(\frac{2\varphi}{f}\right)\right]\right\}\psi\,.
\end{align}

The expression~(\ref{eq:mainferm}) shows that, in the limit $m_\psi\to 0$, the fermionic degree of freedom  decouples from the inflaton. On the other hand, the form~(\ref{eq:lagrY}) of the fermionic Lagrangian emphasizes the shift-symmetric nature of the fermion-inflaton coupling.

As shown in~\cite{Adshead:2018oaa}, the interaction described above, in a quasi-de Sitter background with Hubble parameter $H$, leads to the generation of chiral quanta of $\psi$ with an occupation number that is constant, and given approximately by $.1\left(m_\psi/H\right)^2$, for momenta up to $\sim|\dot\varphi|/f$. The fermions can thus have a very large number density $\sim 10^{-2}\left(\frac{m_\psi}{H}\right)^2\left(\frac{|\dot\varphi|}{f}\right)^3\gg H^3$, and can affect the dynamics of the inflaton background and of its perturbations. In this paper we will be interested in the regime in which the fermions do not affect significantly the background dynamics, but provide the main source of inflaton perturbations.

An especially interesting result of~\cite{Adshead:2018oaa} is that, even in the regime in which the component sourced by the fermions dominates the inflaton perturbations, the statistics of those perturbations is very close to gaussian, and in agreement with the constraints from Planck~\cite{Akrami:2019izv}.  This is due to the fact that, even if the process $\bar\psi\,\psi\to \delta\varphi$ is a $2\to 1$ process that would naturally lead to non gaussian statistics, fermions from a broad set of momenta participate to the process, and gaussianity is re-obtained as an effect of the central limit theorem. The bottom line is that the model of~\cite{Adshead:2018oaa,Adshead:2019aac} can lead to a regime where the perturbations are sourced by the fermions, and still their properties are in agreement with observations.

Since the amplitude of the sourced perturbations has a functional dependence on the parameters of the system that is different from the standard case, this set up has the potential of reviving models of inflation whose potential would be otherwise ruled out by CMB constraints.

We will focus here on the model of inflation where the potential has the simplest functional form: a quadratic potential. In the standard case in which the perturbations are from  the vacuum, this model's prediction for the spectral index is in agreement with data, but is ruled out by the amplitude of the tensor modes, since it predicts $r=8/N$, where $N$ is the number of efoldings, that for $N\lesssim 60$ requires $r>.13$, whereas Planck/Keck constrains $r<.06$.

The presence of fermions in the dynamics in the system naturally calls for a supersymmetric construction. In the next section we will construct a globally supersymmetric model where we obtain the desired features, before moving on to a construction in supergravity that is more complicated, but more appropriate, since we are discussing large field inflation.

\section{A model in global supersymmetry}%
\label{sec:susy}

As a warm up, let us consider a globally supersymmetric theory with a single chiral superfield $\Phi$ and superpotential
\begin{align}\label{eq:w_globalsusy}
W=\frac{\mu}{2}\,\Phi^2+\Lambda^3\,e^{-\sqrt{2}\Phi/F}\,,
\end{align}
where $\mu$, $\Lambda$ and $F$ are parameters with dimensions of mass. The corresponding  Lagrangian reads 
\begin{align}
{\cal L}=-\partial_\mu\phi\,\partial^\mu\phi^*-\left|\mu\phi-\sqrt{2}\frac{\Lambda^3}{F}e^{-\sqrt{2}\phi/F}\right|^2+\bar{\psi}\left[i\gamma^\mu\partial_\mu-\mu-\Re\left\{2\frac{\Lambda^3}{F^2}e^{-\sqrt{2}\phi/F}\right\}+i\Im\left\{2\frac{\Lambda^3}{F^2}e^{-\sqrt{2}\phi/F}\right\}\gamma^5\right]\psi\,,
\end{align}
where $\phi$ is a complex scalar and $\psi$ is a four-component Majorana fermion.

To proceed, we assume\footnote{This is by no means a consistent assumption, and we will make it in this section that has only illustrative purposes. In Section~\ref{sec:sugra} below, on the other hand, we will consistently minimize the full potential of the model.} that the real part of the field $\phi$ is stabilized to $\Re\{\phi\}=0$ and we thus redefine $\phi=i\varphi/\sqrt{2}$, obtaining our final Lagrangian
\begin{align}\label{eq:lagr}
&{\cal L}=-\frac{1}{2}\partial_\mu\varphi\,\partial^\mu\varphi-V(\varphi)+\bar{\psi}\left(i\gamma^\mu\partial_\mu-\mu-2\frac{\Lambda^3}{F^2}\cos(\varphi/F)-2\frac{\Lambda^3}{F^2}i\gamma^5\sin(\varphi/F)\right)\psi\,,\nonumber\\
&V(\varphi)=\frac{\mu^2}{2}\varphi^2-2\,\mu\frac{\Lambda^3}{F}\varphi\,\sin(\varphi/F)+2\frac{\Lambda^6}{F^2}\,.
\end{align}
The fermionic part of this Lagrangian is analogous, with the identifications $F=f/2$ and $m_\psi=2\,\Lambda^3/F^2$, to the Lagrangian~(\ref{eq:mainferm}), with the addition of a mass term $\mu$ for the fermions (that, as we will see below, can be neglected), while, neglecting the cosmological constant $\sim \Lambda^6/F^2$ that we assume to be renormalized to its observed value by some mechanism, the scalar potential is  that of chaotic inflation with oscillating corrections. 

We thus see that the simple superpotential eq.~(\ref{eq:w_globalsusy}) can already lead to the kind of system outlined in Section~\ref{sec:general} above: a model of quadratic inflation (with small corrections) with a sizable coupling of fermions to the inflaton, where the power spectrum of scalar perturbations may be dominated, in some region of parameter space, by the fermion production and might thus be in agreement with Planck constraints. To make sure that this is the case, however, we must explore the constraints on the parameter space available to the system. 

We will assume that the potential is dominated by its quadratic part, and that fermions give a negligible contribution to the background dynamics, so that all the results from chaotic inflation will carry over. In particular, we will have the approximate slow-roll relations
\begin{align}\label{eq:slowroll}
\dot\varphi\simeq -\sqrt{\frac23}\mu M_P\,,\qquad \varphi\simeq2\,M_P\sqrt{N}\,,\qquad H\simeq \mu\sqrt{\frac{2}{3}N}\,,
\end{align}
where $N\simeq 60$ is the number of efoldings until the end of inflation.

The strength of fermion production is measured by the dimensionless parameter $\xi$, that takes the value
\begin{align}
\xi\equiv\frac{|\dot\varphi|}{4FH}=\frac{1}{4\sqrt{N}}\frac{M_P}{F}\,.
\end{align}
We will assume, as is expected to be the case in UV-complete theories of gravity~\cite{Banks:2003sx,ArkaniHamed:2006dz}, that the parameter $F$ is sub-Planckian, and small enough that $\xi\gtrsim 1$. 

Due to the presence of the term proportional to $\mu$ in the fermionic sector of the Lagrangian, the present system is different, as discussed above,  from that of~\cite{Adshead:2018oaa,Adshead:2019aac}. As shown in~\cite{Adshead:2018oaa}, however, fermion production happens for momenta up to $k_{\rm cutoff}\simeq 2H\xi$. As a consequence, since slow roll requires $\mu\ll H$ while $k_{\rm cutoff}\gtrsim H$, the effect of the parameter $\mu$ does not affect the dynamics of fermions to any significant level, and we can safely neglect it.

Let us now list the constraints on our parameter space.

\begin{itemize}

\item {\em Monotonicity of potential.} In order for the oscillating term in the potential not to spoil the monotonicity of the quadratic part during inflation we require
\begin{align}\label{eq:monotonic}
2\frac{\Lambda^3}{F^2\mu}<1\,.
\end{align}

\item {\em Backreaction.} One can neglect the backreaction of the fermions on the inflating background provided the condition ${m^2_\psi\,\xi}\ll 3\pi\,F^2$ is satisfied~\cite{Adshead:2018oaa}. In our model, this corresponds to the condition
\begin{align}
\frac{1}{3\pi\sqrt{N}}\frac{\Lambda^6\,M_P}{F^7}\lesssim 1\,.
\end{align}

\item {\em Validity of effective field theory.} The term proportional to $\sin(\varphi/F)$ in the scalar potential generates oscillations of frequency $\omega=\dot\varphi/F$ in the Hubble parameter. By requiring that physics occurs at scales below the cutoff $4\pi F$ of the axionic effective field theory, we obtain the constraint $\omega\lesssim4\pi F$, that translates into
\begin{align}\label{eq:eft}
\mu M_P\lesssim 2\sqrt{6}\,\pi F^2\,.
\end{align}

\item {\em Small $m_\psi$ approximation.} The results of~\cite{Adshead:2018oaa,Adshead:2019aac} have been obtained assuming $m_\psi\ll H$, that in terms of our parameters reads
\begin{align}\label{eq:smallmu}
\frac{\Lambda^3}{F^{2}}\ll \mu\,\sqrt{\frac{N}{6}}\,,
\end{align}
that is identically satisfied when the condition of monotonicity of the potential, eq.~(\ref{eq:monotonic}) is satisfied. Also, eq.~(\ref{eq:smallmu}), along with the requirement $\xi\gg 1$, implies that a second condition of perturbativity required in~\cite{Adshead:2018oaa,Adshead:2019aac}, namely that $m_\psi\ll H\sqrt{\xi}$, is identically satisfied.  

\item {\em Tensor modes.} The Planck-Keck constraint~\cite{Ade:2018gkx} $r<.06$ rules out standard quadratic inflation. As discussed in~\cite{Adshead:2019aac}, the amplitude of tensor modes in our model has essentially the same expression as in the standard case, which gives a constraint
\begin{align}\label{eq:rconstr}
\frac{4}{3\pi^2}\frac{\mu^2}{M_P^2}N<.06\, P_\zeta\,,
\end{align}
where $P_\zeta$ is the scalar power spectrum.

\item {\em No oscillations in scalar perturbations.} Oscillations in the potential will induce oscillations in the power spectrum of scalar perturbations. This phenomenon has been studied in detail in the case in which the scalar spectrum is generated by the standard mechanism of amplification of vacuum fluctuations of the inflaton. The constraint on the amplitude of those oscillations, $\delta n_s$, can be approximately written as $\delta n_s\lesssim 10^{-3}\sqrt{M_P/F}$~\cite{Easther:2013kla,Akrami:2018odb}, where in our model $\delta n_s\simeq 6\sqrt{2\pi}\,N^{1/4}\frac{\Lambda^3}{\mu F^2}\sqrt{F/M_P}$~\cite{Flauger:2010ja}, which would provide a strong additional constraint on our model. However, these constraints do not hold in the regime we will be interested in, where the scalar perturbations are sourced by the fermion field. As a consequence, we will not consider them in our analysis.

\item {\em Nongaussianities.} There are two potential sources of nongaussianities. First, those induced by the presence of the fermion bath in interaction with the inflaton, that have been shown in~\cite{Adshead:2018oaa} to be negligible. Second, there is a possibility of resonant nongaussianities~\cite{Flauger:2010ja} induced by the small oscillations in the inflaton potential. As in the point above, however, the existing estimates of the amplitude of this effect do not hold in the regime of sourced perturbations we are interested in, and we will ignore them here.

\end{itemize}

\begin{figure}[t]
\centering
\includegraphics[scale=.17]{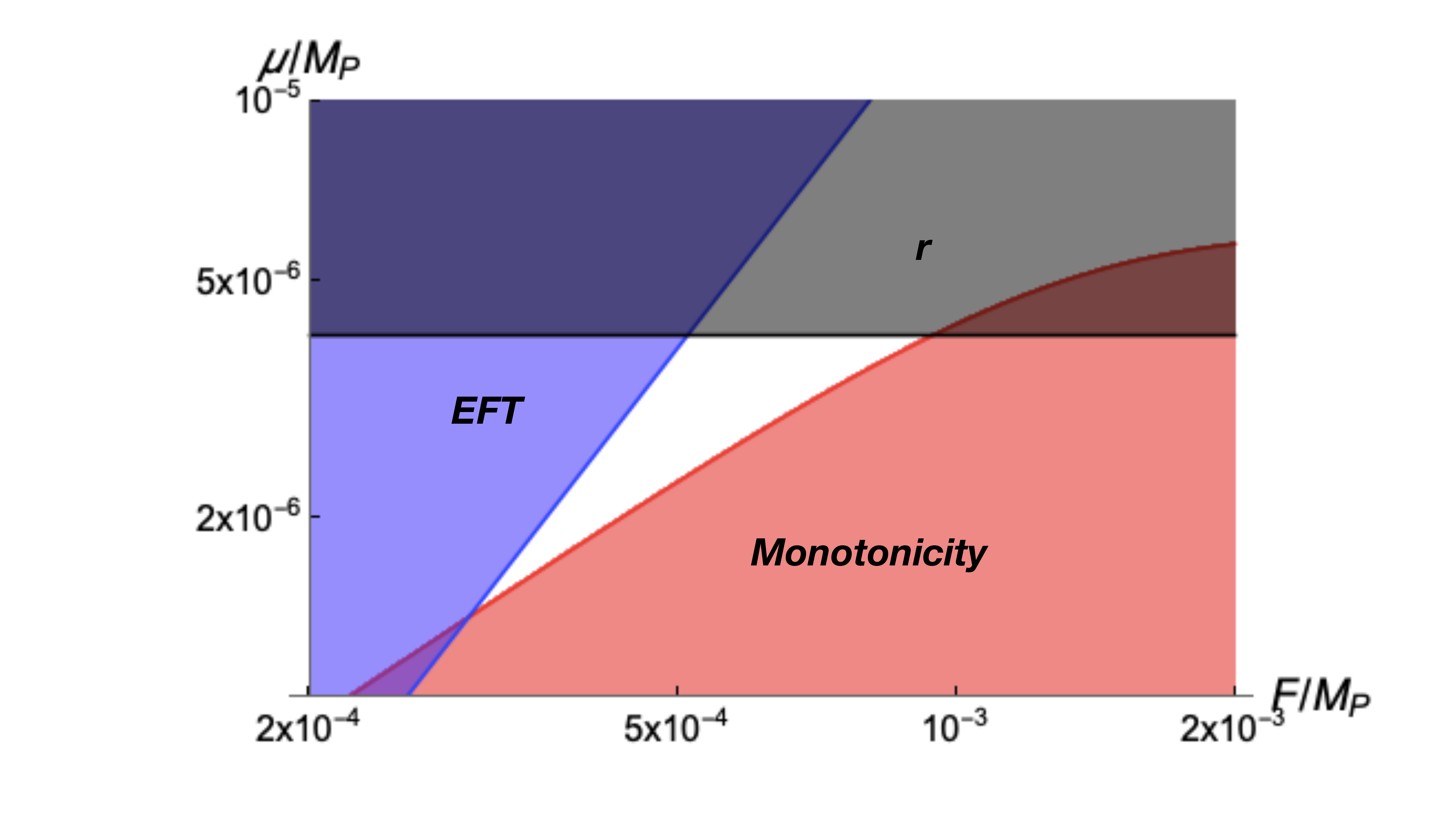} 
\caption{The parameter space for the model (\ref{eq:lagr}). The shaded region marked by an ``$r$'' is excluded by the observational bound $r<.06$ on the amplitude of the tensor modes. The region labelled ``EFT'' is excluded by the constraint~(\ref{eq:eft}), and the region labeled ``Monotonicity'' is excluded by the constraint~(\ref{eq:monotonic}).}
\label{fig:susy}
\end{figure}

Once we fix the number of efoldings to $N=60$, our theory has three parameters, namely $\Lambda$, $\mu$ and $F$. We can eliminate one of them by imposing that the power spectrum takes its observed value $P_\zeta=2.2\times 10^{-9}$, using the expression, obtained in~\cite{Adshead:2018oaa}, 
\begin{align}
P_\zeta=\frac{H^4}{4\pi^2\dot\varphi^2}\left(1+\frac{8}{3\pi^2}\frac{m_\psi^2}{F^2}\xi^2 N\log\xi\right)=\frac{\mu^2\,N^2}{6\pi^2\,M_P^2}\left[1+\frac{2}{3\pi^2}\frac{\Lambda^6\,M_P^2}{F^8}\left|\log\left(4\sqrt{N}\frac{F}{M_P}\right)\right|\right].
\end{align}
In particular, it is convenient to use the normalization of the power spectrum to eliminate $\Lambda$. Once we do this, we can plot the constraints enumerated above on a two-dimensional plot, see Figure~\ref{fig:susy}. As one can see, there is a portion of parameter space that satisfies all the constraints above, and that extends from $F\simeq 3\times 10^{-4}\,M_P$ to $F\simeq 10^{-3}\,M_P$ and where $\mu$ can be as small as $1.3\times 10^{-6}\,M_P$. This implies, using eq.~(\ref{eq:rconstr}), that the tensor-to-scalar ratio in this model of quadratic inflation with corrections can be as small as $\sim .007$, i.e. about an order of magnitude below the present bounds.

Another well-constrained quantity we have not talked about is the spectral index, $.961\lesssim n_s\lesssim.969$~\cite{Aghanim:2018eyx}. Its expression for this specific model is essentially the same as the standard chaotic inflation scenario, $n_s-1=d\log P_\zeta/dN\simeq 2/N$ so that by assuming the standard value $N=60$ the spectral index is automatically in agreement with observations.

To summarize this section, the globally supersymmetric model with superpotential~(\ref{eq:w_globalsusy}), with the  assumption that the real component of the inflaton is stabilized, leads to a model of quadratic inflation that, thanks to inflaton-inflatino interactions, is compatible with all the existing phenomenological constraints. 

Of course, this model with global supersymmetry is not quite suitable for chaotic inflation, where the fields can get Planckian values. In the next section we will thus turn our attention to the more appropriate construction of a model of supergravity where fermions can source the spectrum of scalar perturbations.

\section{The full construction in supergravity }%
\label{sec:sugra}

Even before worrying about the role of fermions, the construction of models of inflation in supergravity is famously~\cite{Dine:1995uk} a nontrivial task. In this paper we will consider models with a stabilizer~\cite{Kawasaki:2000yn,Kallosh:2010ug,Kallosh:2010xz}, that allow to design essentially any potential. The down side of these models is that they need at least two superfields -- the inflaton and the stabilizer, which makes the analysis of the fermionic sector quite cumbersome. 

In Section~\ref{subsec:diagonalize} below, we will study in general terms the equations of motion for the fermionic degrees of freedom in models with an inflaton and a stabilizer superfields. Then, in Section~\ref{subsec:parameters}, we will specialize our equations to the case of a superpotential leading to quadratic inflation with small oscillations, and we will show that the analysis of parameter space performed in the globally supersymmetric model of Section~\ref{sec:susy} above can be directly applied to the full supergravity construction.

\subsection{Equations for fermions in models of supergravity with a stabilizer}%
\label{subsec:diagonalize}

We start from a theory of two chiral multiplets coupling to the supergravity multiplet. 
Of the two spin-1/2 matter fields, one is the goldstino and can be gauged away. We are thus left with two helicity-$1/2$ fermions, the transverse component of the gravitino, $\theta = \gamma^i\psi_i$~\cite{Kallosh:1999jj,Giudice:1999yt,Giudice:1999am}, and the  fermion $\Upsilon$~\cite{Kallosh:2000ve}, a linear combination of the fermions in the matter multiplets. The longitudinal, helicity-$3/2$ component of the gravitino will play no significant role (it gets a mass proportional to the superpotential~\cite{Kallosh:1999jj,Giudice:1999yt,Giudice:1999am}, which vanishes in the models with stabilizer we are interested in), and we will ignore it here. The derivation of the equations of motions for fermions in general supergravity models can be found in~\cite{Kallosh:2000ve}, whose convention we adopt. In particular, eq. (9.20) in that paper provides the equations of motion for the fermions:
\begin{align}
&(\hat{\partial}_0 + \hat{B} + i\gamma^i k_i\gamma^0 \hat{A})\,\theta - \frac{4}{\alpha a}k^2\Upsilon  = 0\,, \nonumber
\\
&(\hat{\partial}_0 - i\gamma^ik_i\gamma^0\hat{A} + \hat{B}^\dagger + a\hat{F} + 2\dot{a} + \frac{a}{M_P^2}\mathbf{m}\gamma^0)\Upsilon + \frac{1}{4}a\alpha\Delta^2\theta = 0\,, \label{eq:eom-original}
\end{align}
where, for a K\"ahler potential $K$ and a superpotential $W$, and considering the two superfields $\Phi_i$, $i=1,2$, and their scalar components $\phi_i$ and $\phi^i=(\phi_i)^*$, one has the quantities
\begingroup
\allowdisplaybreaks
\begin{alignat}{3} \label{eq:formulae}
&m=\mathrm{e}^{\frac{K}{2M_P^2}}W\,, && \mathbf{m} = \Re\{m\} - i\,\Im\{m\} \,\gamma^5\,,&& \nonumber\\
&m^i = \Big(\partial^i + \frac{1}{2M_P^2}\partial^iK\Big)\,m\,,&&  m^{ij} = \Big(\partial^i + \frac{1}{2M_P^2}\partial^iK\Big)\,m^j - \Gamma^{ij}_k\,m^k\,,&&\nonumber\\
&\hat{\partial}_0 = \partial_0 - \frac{i}{2}A_0^B\gamma^5\,, && A_0^B = \frac{i}{2M_P^2}(\phi'{}^i\partial_iK - \phi'_i\partial^iK)\,,&& \nonumber
\\
&H^2=\frac{1}{3M_P^2}\left(|\dot\phi|^2+V\right)\,,&& V=m_i(g^{-1})^i_jm^j-3\frac{|m|^2}{M_P^2}\,,&& |\dot\phi|^2\equiv g^i_j\,\dot\phi_i\,\dot\phi^j\,,\nonumber
\\
&\alpha = 3M_P^2\Big(H^2 + \frac{|m|^2}{M_P^4}\Big)\,, && \alpha_1 = -3M_P^2\Big(H^2 + \frac{2}{3}\dot{H} + \frac{|m|^2}{M_P^4}\Big)\,,&& \alpha_2 = 2\dot{\mathbf{m}}^\dagger\,,\nonumber
\\
&\hat{A} = \frac{1}{\alpha}(\alpha_1 - \gamma^0\alpha_2)\,,&&  \hat{B} = -\frac{3}{2}\dot{a}\hat{A} + \frac{a}{2M_P^2}\mathbf{m}\gamma^0(1+3\hat{A})\,,&&\nonumber
\\
&\xi_i = m_i - \gamma^0g_i^j\dot\phi_j\,,&& \Delta = 2\frac{\sqrt{V}|\dot{\phi}|}{\alpha}\,,&& \nonumber
\\
&P_R=\frac{1}{2}(1-\gamma^5)\,,&& P_L=\frac{1}{2}(1+\gamma^5)\,,&& \Pi_{ij}=\frac{1}{\alpha}(m_ig^k_j\dot\phi_k - m_jg^k_i\dot\phi_k)\,,\nonumber
\\
& \hat{F} = -\frac{4}{\alpha\Delta^2\det{g}}\Big(\xi^kP_R\,(g^{-1})^l_km_{li}&&\Pi^{ij}\xi_j^\dagger + \xi_kP_L(g^{-1})^k_lm^{li}\Pi_{ij}\xi^{\dagger j}\Big)\,,&&
\end{alignat}
\endgroup
where a prime denotes a derivative with respect to the conformal time while an overdot is a derivative with respect to cosmic time, and where $\partial^i = \partial/\partial\phi_i$, and $\partial_i = \partial/\partial\phi^i$. Also, $g^i_j\equiv\partial^i\partial_jK$ is the K\"ahler metric and $\Gamma^{ij}{}_k =(g^l{}_k)^{-1}\partial^ig^j{}_l$ is the K\"ahler connection. The scalars satisfy the equations of motion $g^j_i(\ddot{\phi}_j + 3H\dot{\phi}_j + \Gamma^{kl}_j\dot{\phi}_k\dot{\phi}_l) + \partial_iV = 0$.
Further, we have the relation $\alpha_1^2 + \alpha_2^\dagger \alpha_2+ \alpha^2\Delta^2 = 1$.

We will denote the two chiral superfields by $\Phi = \Phi_1$, and $S=\Phi_2$ (with scalar components $\phi$ and $s$, respectively), and  we choose a minimal K\"ahler potential for $S$, but keep a general potential for $\Phi$,
\begin{align}\label{eq:gen_kae}
K(\Phi,\,\bar\Phi; S,\,\bar{S}) = \mathcal{K}(\Phi,\,\bar\Phi) + S\bar{S}\,.
\end{align}

For the superpotential, we use a stabilizer model,
\begin{align}\label{eq:stab_W}
W = S\,f(\Phi)\,,
\end{align}
where $f(\Phi)$ is an arbitrary function and $S$ is stabilized at 0. A consequence of this is that $m|_{s=0}=0$, and therefore the mass of the longitudinal helicity-3/2 component of the gravitino, $m_{3/2} = |m|/M_P^2$ vanishes.

The scalar potential is
\begin{align}
V &= \mathrm{e}^{\frac{K}{M_P^2}}\left[(g^{-1})^i_j\left(\partial_i{W}^* + \frac{\partial_iK}{M_P^2}{W}^*\right)\left(\partial^jW + \frac{\partial^jK}{M_P^2}W\right) - 3\frac{|W|^2}{M_P^2}\right]\nonumber
\\
&= \mathrm{e}^{\frac{K}{M_P^2}}\Big[\Big|f(\phi) + \frac{|s|^2}{M_P^2}f(\phi)\Big|^2 + g^\phi_\phi|s|^2\Big|f'(\phi) + \frac{\partial^\phi K}{M_P^2}f(\phi)\Big|^2 - 3\frac{|s|^2}{M_P^2}|f(\phi)|^2\Big]\,.
\end{align}

Differentiating gives $\partial_sV|_{s=0} = 0$, $\partial_s^2V|_{s=0} \geq 0$, and   $\partial_s\partial_{\bar{s}} V|_{s=0} \geq 0$ showing that $s=0$ is a stable critical point of the potential. Therefore, from here on we set $s=0$, and the scalar potential is simply $V = \e^\frac{\mathcal{K}}{M_P^2}|f(\phi)|^2$.

With these choices, we have
\begin{alignat}{3}
&m^s = \mathrm{e}^{\frac{\mathcal{K}}{2M_P^2}}f(\phi)\,,\qquad && m^{s\phi} = \mathrm{e}^{\frac{\mathcal{K}}{2M_P^2}}\Big[f'(\phi) + \frac{\partial^\phi {\mathcal{K}}}{M_P^2}f(\phi)\Big]\,,\qquad && m^\phi = m^{ss} = m^{\phi\phi} = 0\,,\\
&\xi^s = \mathrm{e}^{\frac{\mathcal{K}}{2M_P^2}}f(\phi)\,, &&\xi^\phi = -\gamma^0\,g^\phi_\phi\dot{{\phi}}^*\,, && \Pi^{s\phi} =\, \frac{1}{\alpha}\mathrm{e}^{\frac{\mathcal{K}}{2M_P^2}}f(\phi)g^\phi_\phi\dot{{\phi}}^*\,.
\end{alignat}

A bit of calculation shows that
\begin{align}
\hat{F} &= \frac{V-|\dot{\phi}|^2}{2V|\dot{\phi}|^2}\Big(\partial^\phi V\dot{\phi} + \partial_\phi V\dot{{\phi}}^*\Big) + \frac{V+|\dot{\phi}|^2}{2V|\dot\phi|^2}\Big(\partial^\phi V\dot{\phi} - \partial_\phi V\dot{{\phi}}^*\Big)\gamma^5\,.
\end{align}

Let us now proceed to diagonalize the equations of motion for the fermions. The system \eqref{eq:eom-original} can be derived from the Lagrangian \cite{Nilles:2001fg}
\begin{align}
\mathcal{L} &= -\frac{\alpha a^3}{4k^2}\bar{\theta}\Big[\Big( \gamma^0\hat{\partial}_0 + i\gamma^ik_i\hat{A} +\gamma^0\hat{B} \Big)\theta - \frac{4k^2}{a\alpha}\gamma^0\Upsilon\Big]\,+\nonumber
\\
&- \frac{4a}{\alpha\Delta^2}\bar{\Upsilon}\Big[ \Big(\gamma^0\hat{\partial}_0 - i\gamma^ik_i\hat{A} + \gamma^0\hat{B}^\dagger + a\gamma^0\hat{F} + 2\dot{a}\gamma^0 + \frac{a}{M_P^2}\gamma^0\mathbf{m}\gamma^0\Big)\Upsilon + \frac{1}{4}a\alpha\Delta^2\gamma^0\theta \Big], \label{eq:lag-original}
\end{align}
where, following~\cite{Kallosh:2000ve}, we use the convention $\bar{\theta} = i\theta^\dagger\gamma^0$ for barred spinors. We canonically normalize the fermions defining
\begin{align}
\theta = 2\frac{i\gamma^ik_i}{\sqrt{\alpha a^3}}\tilde{\theta}\,, &\hspace{1cm} \Upsilon = \frac{\Delta}{2}\Big(\frac{\alpha}{a}\Big)^{1/2}\tilde{\Upsilon}\,.
\end{align}
The Lagrangian with normalized fields (and taking $s=0$) is
\begin{align}\label{eq:lagr-norm}
\mathcal{L} &= \bar{\tilde{\theta}} \Big[\Big( -\gamma^0\partial_0 + i\gamma^ik_i\frac{\alpha_1}{\alpha} - \frac{i}{2}A_0^B\gamma^0\gamma^5 \Big) \tilde{\theta} + i\Delta\gamma\cdot k\gamma^0 \tilde{\Upsilon} \Big]\,+\nonumber
\\
& + \bar{\tilde{\Upsilon}} \Big[ \Big(-\gamma^0\partial_0 + i\gamma^ik_i\frac{\alpha_1}{\alpha} + \Big(\frac{i}{2}A_0^B - a\hat{F}_5\Big)\gamma^0\gamma^5 \Big)\tilde{\Upsilon} + i\Delta\gamma\cdot k\gamma^0\tilde{\theta} \Big]\,.
\end{align}
where
\begin{align} \label{eq:f5-exact}
\hat{F}_5 &= \frac{V+|\dot{\phi}|^2}{2V|\dot\phi|^2}\Big(\partial^\phi V\dot{\phi} - \partial_\phi V\dot{{\phi}}^*\Big)\,.
\end{align}
Note in particular that $\hat{F}_5$ is pure imaginary, and that it vanishes for real $\phi$. 

Let us write the Lagrangian~(\ref{eq:lagr-norm}) in the compact form
\begin{align}
\mathcal{L} &= \bar{X}\Big[-\gamma^0\partial_0 + i\gamma\cdot k\, N + M\Big]X\,,
\end{align}
with $X = (\tilde{\theta}, \tilde{\Upsilon})^T$ and $N = N_1 + N_2\gamma^0$, where
\begin{align}
N_1 = 
\begin{pmatrix}
\alpha_1/\alpha & 0 \\
0 & \alpha_1/\alpha
\end{pmatrix},
\hspace{1cm}
N_2 = 
\begin{pmatrix}
0 & \Delta \\
\Delta & 0
\end{pmatrix},
\hspace{1cm}
M = 
\begin{pmatrix}
-\frac{i}{2}A_0^B & 0 \\
0 & \frac{i}{2}A_0^B - a\hat{F}_5
\end{pmatrix}
\gamma^0\gamma^5\,.
\end{align}

We now redefine the fields in such a way as to remove the factor of $N$ in front of $i\gamma^ik_i$. Using the relation $\alpha^2 - \alpha_1^2 = \alpha^2\Delta^2$, we can see that $N^\dagger N = N_1^2+N_2^2=1$, so $N$ is unitary. Therefore, we can define $N = \mathrm{e}^{2\Psi\gamma^0} = \cos2\Psi + \gamma^0\sin2\Psi$ where $\Psi$ is a $2\times2$ hermitian matrix~\cite{Peloso:2000hy}. We choose
\begin{align}
2\Psi = 
\begin{pmatrix}
0 & \pi-\sin^{-1}\Delta \\
\pi-\sin^{-1}\Delta & 0
\end{pmatrix}.
\end{align}
It is straightforward to check that $\cos2\Psi = N_1$ (remember that $\alpha_1<0$), and $\sin2\Psi = N_2$. After redefining $X = \mathrm{e}^{-\Psi\gamma^0}Z$, the Lagrangian takes the form
\begin{align}
\mathcal{L} &= \bar{Z}\Big[-\gamma^0\partial_0 + i\gamma\cdot k + \tilde{M}\Big]Z\,,
\end{align}
where the new matrix $\tilde{M}$ reads
\begin{align}
\tilde{M} &\equiv \mathrm{e}^{\Psi\gamma^0}(M-\partial_0\Psi)\,\mathrm{e}^{-\Psi\gamma^0}\nonumber
\\
&= \frac12\begin{pmatrix}
\big[-iA_0^B + a\,(1-\alpha_1/\alpha)\,\hat{F}_5\big]\gamma^0\gamma^5 & -\frac{\alpha}{\alpha_1}\Delta' + a\,\hat{F}_5\Delta\gamma^5
\\
-\frac{\alpha}{\alpha_1}\Delta' + a\,\hat{F}_5\Delta\gamma^5 & \big[iA_0^B - a\,(1+\alpha_1/\alpha)\,\hat{F}_5\big]\gamma^0\gamma^5
\end{pmatrix}.
\end{align}
Furthermore, we can remove the $\gamma^0\gamma^5$ term by redefining the fields as
\begin{align}
Z=\begin{pmatrix}
\mathrm{e}^{i\sigma_1\gamma^5} & 0 \\
0 & \mathrm{e}^{i\sigma_2\gamma^5}
\end{pmatrix}
\begin{pmatrix}
\psi_1 \\
\psi_2
\end{pmatrix},
\end{align}
where, in order for the $\gamma^0\gamma^5$ terms to vanish, $\sigma_1$ and $\sigma_2$ must satisfy
\begin{align} 
\sigma_1' &= -\frac{1}{2}A_0^B -i\frac{a}{2}\,(1-\alpha_1/\alpha)\,\hat{F}_5\,,\nonumber
\\
\sigma_2' &= \frac{1}{2}A_0^B +i \frac{a}{2}\,(1+\alpha_1/\alpha)\,\hat{F}_5\,.
\end{align}

Once we choose $\sigma_1$ and $\sigma_2$ that satisfy these equations, we are at last left with a coupled set of fermions with a mass matrix of the form $\begin{pmatrix}0&M_1+iM_2\gamma^5\\M_1+iM_2\gamma^5&0\end{pmatrix}$, where $M_1$ and $M_2$ are defined below. Such a system can be completely diagonalized in terms of the rotated fields
\begin{align}
\chi_1 &= \frac{1}{\sqrt{2}}(\psi_1 + \psi_2)\,,\nonumber
\\
\chi_2 &= \frac{1}{\sqrt{2}}(\psi_1 - \psi_2)\,,
\end{align}
giving the final Lagrangian
\begin{align}
\mathcal{L} &= (\bar{\chi}_1, \bar{\chi}_2) \left[-\gamma^0\partial_0 + i\gamma\cdot k + a\begin{pmatrix}
M_1+iM_2\gamma^5 & 0
\\
0 & -M_1-iM_2\gamma^5
\end{pmatrix} \right]
\begin{pmatrix}
\chi_1 \\
\chi_2
\end{pmatrix}\,, \label{eq:final-exact-lagrangian}
\end{align}
where
\begin{align}\label{eq:final-mass}
M_1 &= -\frac{\alpha}{2\alpha_1}\dot{\Delta}\cos(\sigma_1+\sigma_2) + \frac{i}{2}\hat{F}_5\Delta\sin(\sigma_1+\sigma_2)\,,\nonumber
\\
M_2 &= -\frac{\alpha}{2\alpha_1}\dot{\Delta}\sin(\sigma_1+\sigma_2) - \frac{i}{2}\hat{F}_5\Delta\cos(\sigma_1+\sigma_2)\,,
\end{align}
that depend only on the combination
\begin{align}\label{eq:sigma-exact}
\sigma\equiv \sigma_1+\sigma_2\,,\qquad \dot{\sigma} = i\frac{\alpha_1}{\alpha}\hat{F}_5\,,
\end{align}
and where $\alpha$, $\alpha_1$, and $\Delta$ are given in~\eqref{eq:formulae}, and $\hat{F}_5$ is given in eq.~\eqref{eq:f5-exact}. Thus, we see that we have a system of two decoupled fermions with the same mass. This is a general result, assuming only a stabilizer model superpotential where the K\"ahler potential is  minimal in $S$. The scalar potential and fermion dynamics are determined by the choice of function $f(\Phi)$ and K\"ahler potential, ${\mathcal K}(\Phi,\,\bar\Phi)$. This allows a great deal of freedom in constructing a model with fermions coupled to an inflaton with choice of inflationary potential. For example, taking $\phi$ to be real will make $M_2 = 0$ and $M_1 = -\frac{\alpha}{2\,\alpha_1}\Delta$.

In the next section, we show how this can be used to recover, in a full supergravity setting, the Lagrangian of Section~\ref{sec:susy}.

\subsection{Quadratic inflaton potential, plus small oscillations -- analysis of the parameter space}%
\label{subsec:parameters}

We now show how we can recover the Lagrangian~\eqref{eq:lagr} from the full supergravity theory in \eqref{eq:final-exact-lagrangian} with the choice
\begin{align}\label{eq:sugra_WK}
f(\Phi) &= \mu\Phi + \hat{\Lambda}^2\mathrm{e}^{-\frac{\sqrt{2}\Phi}{F}}\,,\hspace{1cm}{\mathcal K}(\Phi,\,\bar\Phi) = \frac{1}{2}(\Phi+\bar{\Phi})^2\,.
\end{align}

We have three parameters, $\mu$, $F$, and $\hat{\Lambda}$ with the dimensions of mass. Here, we write $\hat{\Lambda}$ to distinguish the parameter of this section from the $\Lambda$ of Section \ref{sec:susy}. We take $\phi = \frac{1}{\sqrt2}(\rho + i\varphi)$ so that the scalars are canonically normalized. During inflation, $\varphi$ will act as the inflaton while $\rho$ will oscillate near its minimum and will not play a significant role in the scalar potential. The choice~(\ref{eq:sugra_WK}) gives the scalar potential
\begin{align}\label{eq:sugra_potential}
V &= \mathrm{e}^{\frac{\rho^2}{M_P^2}}\Big[ \frac{\mu^2}{2}(\rho^2+\varphi^2) + \sqrt{2}\mu\,\hat{\Lambda}^2\mathrm{e}^{-\frac{\rho}{F}}\Big(\rho\cos\frac{\varphi}{F} - \varphi\sin\frac{\varphi}{F}\Big) + \hat{\Lambda}^4\mathrm{e}^{-\frac{2\rho}{F}} \Big]\,.
\end{align}

We will take there to be a hierarchy of scales, $\rho \ll F \ll M_P \lesssim \varphi$. As we will see below, therefore, $\rho$ will be nonzero, but can be made sufficiently small within a certain parameter range. As mentioned in Section~\ref{sec:susy}, $F\ll M_P$ is motivated by embedding this model in a UV-complete theory of gravity. The scalar potential is then well approximated by
\begin{align}
V &\simeq \frac{\mu^2}{2}\varphi^2 - \sqrt{2}\mu\hat{\Lambda}^2\varphi\sin\frac{\varphi}{F} + \hat{\Lambda}^4\,.
\end{align}

This potential has the same for as the one given in eq.~\eqref{eq:lagr}, namely chaotic inflation plus small oscillations. Matching gives the relation 
\begin{align}\label{eq:match}
\hat{\Lambda}^2 = \sqrt{2}\Lambda^3/F\,,
\end{align}
so that monotonicity of the potential requires
\begin{align} \label{eq:monotonic-2}
\sqrt{2}\frac{\hat{\Lambda}^2}{\mu F} < 1\,.
\end{align}
Once this condition is satisfied, we can use the slow-roll approximation~(\ref{eq:slowroll}) to describe the evolution of $\varphi$ at zeroth order.

We can now solve for $\rho(t)$ from the equations of motion obtained after keeping only the leading terms in the potential~(\ref{eq:sugra_potential}) under the hierarchy $\rho \ll F \ll M_P \lesssim \varphi$,
\begin{align}\label{eq:eq_rho}
\ddot{\rho} + 3H\dot{\rho} + \frac{\sqrt{2}\mu\hat{\Lambda}^2}{F} \varphi\sin\frac{\varphi}{F} = 0\,,
\end{align}
where we will treat $\varphi$ as approximately constant except in the rapidly oscillating $\sin(\varphi/F)$ term where we use the leading order in slow-roll $\varphi(t) \simeq \varphi(0)-\sqrt{\frac{2}{3}}\mu M_Pt$. Neglecting the decaying term from the homogeneous solution of eq.~(\ref{eq:eq_rho}), and requiring 
\begin{align}
\frac{F}{M_P} \ll \frac{M_P}{\varphi}\propto\frac{1}{\sqrt{N}}\,,
\end{align}
that is equivalent to the large-$\xi$ approximation, we obtain 
\begin{align} \label{eq:rho}
\rho(t) &\simeq \rho_0+ \frac{3}{\sqrt{2}}\frac{\hat{\Lambda}^2}{\mu M_P}\frac{F}{M_P}\varphi\sin\frac{\varphi}{F}\,,
\end{align}
where $\rho_0$ is an integration constant. We have verified, by solving numerically the exact system of coupled equations for $\rho$ and $\phi$, the accuracy of the approximation~(\ref{eq:rho}) and that  the constant $\rho_0$ is much smaller than $F$.

The requirement $\rho \ll F$ gives, therefore, the additional constraint
\begin{align} \label{eq:rholessF}
\frac{\hat{\Lambda}^2}{\mu M_P} \ll \frac{M_P}{\varphi} \simeq \frac{1}{\sqrt{N}}\,.
\end{align}

Now we move on to $\varphi(t)$, for which we want to go beyond the slow-roll approximation.  The function $\varphi(t)$ satisfies the approximate equation
\begin{align}
\ddot{\varphi} + 3H\dot{\varphi} + \mu^2\varphi \left(1- \sqrt{2}\frac{\hat{\Lambda}^2}{\mu F}\cos\frac{\varphi}{F}\right) = 0\,,
\end{align}
that we can solve perturbatively in $\hat{\Lambda}^2$, defining $\varphi = \varphi_0 + \hat\Lambda^2\varphi_1 + O(\hat{\Lambda}^4)$ \cite{Flauger:2009ab, Flauger:2010ja}. By linearizing the equation for $\varphi$ in $\hat\Lambda^2$, and keeping the leading terms in the approximation $\varphi\gtrsim M_P\gg F$ and in the slow roll approximation, the equation for $\varphi_1$ reads
\begin{align}
\ddot{\varphi}_1 + \sqrt{\frac{3}{2}}\frac{\mu\varphi_0}{M_P}\dot{\varphi}_1 - \sqrt{2}\frac{\mu\varphi_0}{F}\cos\frac{\varphi_0}{F} = 0\,,
\end{align}
where, again, we  treat $\varphi_0$ as constant except inside the rapidly oscillating $\cos(\varphi_0/F)$ term. The solution, ignoring the decaying mode, is
\begin{align}\label{eq:phi}
\varphi(t) \simeq \varphi_0(t) - \frac{3}{\sqrt{2}}\frac{\hat{\Lambda}^2}{\mu M_P}\frac{F}{M_P}\varphi_0(t)\cos\left(\frac{\varphi_0(t)}{F}\right)\,.
\end{align}
We see from~\eqref{eq:rholessF} and from $F\ll M_P$ that $\hat\Lambda^2\varphi_1 \ll \varphi_0$, therefore, we are comfortably within the perturbative region. 

Now we turn our attention to the remaining quantities in the fermion mass in eq. \eqref{eq:final-exact-lagrangian}, starting with $\hat{F}_5$. Using \eqref{eq:rho},\eqref{eq:phi}, along with the approximations \eqref{eq:rholessF}, this gives

\begin{align} \label{eq:f5}
\hat{F}_5 \simeq \frac{-\sqrt{3}i\hat{\Lambda}^2}{M_P}\frac{\varphi}{F}\sin(\varphi/F)\,.
\end{align}

Note that we are using $\varphi$ and not $\varphi_0$ in the above expression. At the order we are considering, they are equivalent. Continuing with $\sigma$, to leading order in slow-roll, $\alpha_1/\alpha \simeq -1$, so that $\dot{\sigma} \simeq -i\hat{F}_5$. When integrating $\dot{\sigma}$ in eq.~(\ref{eq:sigma-exact}), we will treat $\varphi$ as constant outside of the $\sin(\varphi/F)$. We will not be interested in the constant of integration as it is simply a constant phase in the fermion fields, so that we obtain
\begin{align}
\sigma &\simeq -\frac{3\hat{\Lambda}^2}{\sqrt{2}\mu M_P^2}\varphi\cos(\varphi/F)\,.\, \label{eq:sigma}
\end{align}

Performing the same approximations for $\Delta$, we obtain
\begin{align}
\Delta &\simeq \sqrt{\frac{2}{3}}\frac{2M_P}{\varphi}\Big( 1 - \frac{2M_P^2}{3\varphi^2} + \frac{3}{\sqrt{2}}\frac{\hat{\Lambda}^2}{\mu M_P}\frac{\varphi}{M_P}\sin(\varphi/F)\Big)\,, \label{eq:delta}
\end{align}
and
\begin{align}
\dot{\Delta} &\simeq \frac{4}{3}\frac{\mu M_P^2}{\varphi^2} - 2\sqrt{2}\frac{\hat{\Lambda}^2}{F}\cos(\varphi/F)\,. \label{eq:deltadot}
\end{align}

\begin{figure}
\centering
\includegraphics[scale=.7]{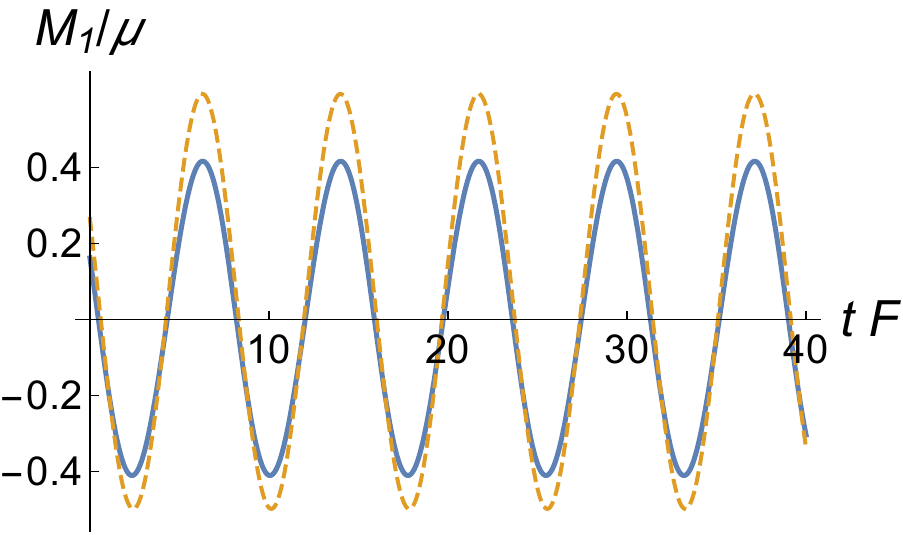} 
\hspace{1cm}
\includegraphics[scale=.7]{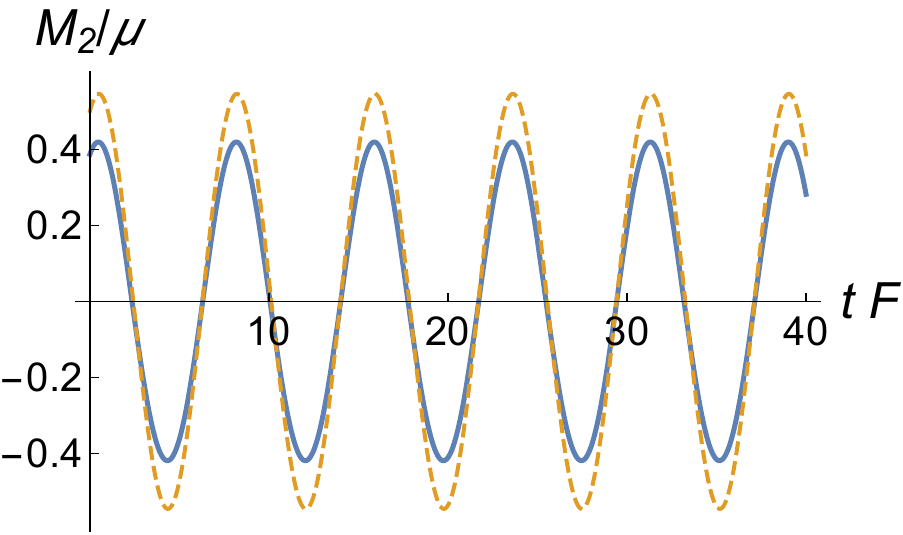} 
\caption{Results of exact numerical integration (solid, blue) and analytical approximations, eq.~(\ref{mass-final-result}), (dashed, orange) for the quantities $M_1(t)$ and $M_2(t)$. The parameters used for these plots are $\mu=5\times10^{-6}\,M_P$, $F=5\times 10^{-4}\,M_P$. At these times $\varphi\simeq 13.9\,M_P$.}
\label{fig:M1M2}
\end{figure}

Finally, inserting \eqref{eq:f5}, \eqref{eq:sigma}, \eqref{eq:delta}, and \eqref{eq:deltadot} into \eqref{eq:final-mass}, we get the fermion mass,
\begin{align}\label{mass-final-result}
M_1 + iM_2\gamma^5 &\simeq \mu\Big(\frac{2M_P^2}{3\varphi^2} - \sqrt{2}\frac{\hat{\Lambda}^2}{\mu F}\big(\cos(\varphi/F) + i\sin(\varphi/F)\gamma^5\big)\Big)\,.
\end{align}
The accuracy of this approximation is shown in Figure~\ref{fig:M1M2}, for a choice of parameters that corresponds approximately to the center of the white region in Figure~\ref{fig:susy}. 

By translating to the parameters of Section~\ref{sec:susy} using the identification~(\ref{eq:match}), we recover the fermionic part of the Lagrangian of equation~\eqref{eq:lagr}. In the supergravity case, the constant part of the fermionic mass (i.e., corresponding to the term proportional to $\mu$ in the first line of eq.~\eqref{eq:lagr}) is slow-roll suppressed, and we can neglect it here as we did in Section~\ref{sec:susy}.

To conclude, with the redefinition~(\ref{eq:match}) the plot in Figure~\ref{fig:susy} applies also to the supergravity model. In particular, this shows that the supergravity model defined by eqs.~(\ref{eq:gen_kae}), (\ref{eq:stab_W}) and~(\ref{eq:sugra_WK}) there is a regime of parameter space where the data can be in agreement with all CMB constraints while the inflaton potential is, up to corrections that we want to be negligible, simply quadratic.

\section{Discussion and conclusions}%
\label{sec:conclusion}

Standard chaotic inflation is ruled out by experiment. It predicts too large a value for the tensor-to-scalar ratio. The tensor spectrum is determined by the energy scale of inflation, which in the simple model of quadratic inflation is fixed by the normalization of the scalar spectrum. We have shown in this paper that a source-dominated scalar spectrum can allow to lower the energy scale of inflation, thereby bringing chaotic inflation back into the observationally allowed regime.

In the papers~\cite{Adshead:2018oaa,Adshead:2019aac} it was shown that fermions coupled to an axion inflaton can lead to a source-dominated scalar spectrum and a vacuum-dominated tensor spectrum. More specifically, since the vacuum perturbations and sourced perturbations of the scalar modes are statistically independent, the power spectrum is the sum, $\mathcal{P}_\zeta = \mathcal{P}_\zeta^{\text{vacuum}} + \mathcal{P}_\zeta^{\text{sourced}}$, and similarly for the tensor spectrum. Therefore, the fermion-sourced model with $2.2\times 10^{-9} \simeq \mathcal{P}_\zeta^{\text{sourced}} \gg \mathcal{P}_\zeta^{\text{vacuum}} \propto V\propto \mathcal{P}_t$, allows one to lower the energy scale of inflation. With $\mathcal{P}_t$ dominated by the vacuum perturbations one can then lower the value of the tensor-to-scalar ratio. 

This work contains two main results. First, we have shown that the model of~\cite{Adshead:2018oaa,Adshead:2019aac} can be effectively constructed from a globally supersymmetric model with superpotential~\eqref{eq:w_globalsusy}. This superpotential generates a quadratic scalar potential, plus small oscillations.  The fermion sector produces the inflaton-fermion coupling studied in~\cite{Adshead:2018oaa,Adshead:2019aac} with a negligible additional fermion mass term. In particular, this applies naturally to the model of~\cite{Kaloper:2008qs,Kaloper:2008fb,Kaloper:2011jz}, that naturally leads to a quadratic inflaton potential using monodromy. Thus, the analysis from~\cite{Adshead:2018oaa,Adshead:2019aac} applies, allowing for the lowering of $r$ while maintaining $n_s$ unaffected and without generating large non-Gaussianities. While the model is subject to a number of constraints, there is a region, in white in Figure~\ref{fig:susy}, where those constraints are all satisfied.

Second, we have examined supergravity with two chiral multiplets with one of the scalars acting as a stabilizer. In Section~\ref{subsec:diagonalize} we have written down the general equations of motions for the fermions in this class of models. Remarkably, the two helicity-$1/2$ states in the theory behave identically, as fermions with mass $M_1 + iM_2\gamma^5$, where the generally time dependent terms $M_1$ and $M_2$ are given in eq.~(\ref{eq:final-mass}). Specializing to the case where the superpotential consists of a slowly varying component and quickly oscillating term, we have shown in Section~\ref{subsec:parameters} that the equations for the fermions in the full supergravity theory reduce to those obtained in the case of the globally supersymmetric model, in agreement with the intuition from the equivalence theorem~\cite{Fayet:1986zc,Casalbuoni:1988kv,Casalbuoni:1988qd}. It would be interesting to see whether these results extend to more general classes, beyond those with a stabilizer, of models of axion inflation in supergravity. It is also worth stressing that, while we have focused on quadratic potential, our construction can be extended to general models of monomial inflation. The bottom line is that, in a class of relatively simple models of inflation in supergravity, the potential can be essentially quadratic while the theory is compatible with all existing observations.

\acknowledgments 

We thank Nemanja Kaloper and Lauren Pearce for very useful discussions. This work is partially supported by the US-NSF grant PHY-1820675.

\end{document}